\documentclass[sigconf]{acmart}

\usepackage{listings}

\usepackage{tabularx}
\usepackage{float}

\usepackage{booktabs}
\usepackage{longtable}
\usepackage{xspace}
\usepackage{graphicx}
\usepackage{textcomp}
\usepackage{algorithm}
\usepackage{algpseudocode}
\usepackage{xcolor}
\usepackage{url}
\usepackage{enumitem}
\usepackage{comment}
\usepackage{array}

\usepackage{multirow}
\usepackage{makecell}
\usepackage{afterpage}
\usepackage{float}
\usepackage{graphicx}
\usepackage{tabularx}
\usepackage{hyperref}
\usepackage{longtable}
\usepackage{multirow}
\usepackage{array}
\usepackage[justification=centering]{caption}
\usepackage{tikz}
\usetikzlibrary{positioning}
\usepackage{soul}
\usepackage{hyperref}
\usepackage{tablefootnote}

\definecolor{source}{gray}{0.95}
\definecolor{highlight}{gray}{0.9}
\definecolor{bblue}{HTML}{4F81BD}
\definecolor{rred}{HTML}{C0504D}
\definecolor{ggreen}{HTML}{9BBB59}
\definecolor{ppurple}{HTML}{9F4C7C}

\usepackage{textcase}  
\usepackage{multirow}

\definecolor{source}{gray}{0.9}

\newcommand{\boxit}[2][gray!15]{%
\vspace{5pt}
  \begin{center}
  \noindent
  \setlength{\fboxsep}{1.5pt}% Set padding around the content
  \setlength{\fboxrule}{0.5pt}% Set border thickness
  \fbox{%
    \colorbox{#1}{%
      \parbox{8.1cm}{#2}%
    }%
  }%
  \end{center}
}

\newcommand\tool{LLMSecGuard\xspace}

\copyrightyear{2024}
\acmYear{2024}
\setcopyright{rightsretained}
\acmConference[EASE 2024]{28th International Conference on Evaluation and Assessment in Software Engineering}{June 18--21, 2024}{Salerno, Italy}
\acmBooktitle{28th International Conference on Evaluation and Assessment in Software Engineering (EASE 2024), June 18--21, 2024, Salerno, Italy}\acmDOI{10.1145/3661167.3661263}
\acmISBN{979-8-4007-1701-7/24/06}

\setcopyright{none}
\acmDOI{10.1145/xxx}
\usepackage{draftwatermark}
\usepackage{transparent}
\SetWatermarkText{\transparent{0.5}\fontsize{70}{74}\selectfont \parbox[c][3cm][c]{\paperwidth}{\centering Preprint Version\\SECUTE 2024}}

\SetWatermarkScale{1}

\begin{document}

\title{LLM Security Guard for Code}

\author{Arya Kavian}
\affiliation{
% Department of Electrical and Computer Engineering,
University of Science and Technology of Mazandaran
\city{Behshahr}
\country{Iran}
}

\author{Mohammad Mehdi Pourhashem Kallehbasti}
\orcid{0000-0001-9484-6813}
\affiliation{
% Department of Electrical and Computer Engineering,
University of Science and Technology of Mazandaran
\city{Behshahr}
\country{Iran}
}

\author{Sajjad Kazemi}
\affiliation{
% Department of Electrical and Computer Engineering,
University of Science and Technology of Mazandaran
\city{Behshahr}
\country{Iran}
}

\author{Ehsan Firouzi}
\orcid{0009-0000-7563-4196}
\affiliation{
Technische Universität Clausthal
\country{Germany}
}

\author{Mohammad Ghafari}
\orcid{0000-0002-1986-9668}
\affiliation{
Technische Universität Clausthal
\country{Germany}
}

\begin{abstract}

Many developers rely on Large Language Models (LLMs) to facilitate software development.
Nevertheless, these models have exhibited limited capabilities in the security domain.
We introduce \tool, a framework to offer enhanced code security through the synergy between static code analyzers and LLMs.
\tool is open source and aims to equip developers with code solutions that are more secure than the code initially generated by LLMs.
This framework also has a benchmarking feature, aimed at providing insights into the evolving security attributes of these models.

\end{abstract}

\keywords{Security analysis, secure code generation, code models}

\begin{CCSXML}
<ccs2012>
   <concept>
       <concept_id>10002978.10003022.10003023</concept_id>
       <concept_desc>Security and privacy~Software security engineering</concept_desc>
       <concept_significance>500</concept_significance>
       </concept>
 </ccs2012>
\end{CCSXML}

\ccsdesc[500]{Security and privacy~Software security engineering}

\maketitle

\section{Introduction}

The use of Large Language Models (LLMs) such as ChatGPT and Copilot has become popular for software development activities such as coding, design, comprehension, etc.~\cite{hou2023large,fan2023large}.
Nevertheless, hallucination, i.e., ``presenting incorrect information as if it is correct'', poses serious challenges for LLM users~\cite{huang2023survey}. 
This issue is more prevalent in domains where reliable training content is scarce, such as in the realm of code security.
Indeed, 
recent studies have shown that code models are widely adopted for code generation~\cite{GitHubBlog}, but they have limited capabilities in the software security domain~\cite{wang2023effectiveness, Asare2023}.
Therefore, a vulnerable code piece that an LLM mistakenly recommends as a secure solution could compromise the entire system's security if it is used without enough scrutiny.

We introduce \tool, a framework specifically designed to examine the security properties of LLMs for code generation and harness their analytical capabilities to facilitate secure code development.
It applies static security analysis on LLM-generated code to uncover potential security issues and guides LLMs in resolving such issues in the code. 
Moreover, \tool can assess the security properties of LLMs and benchmark them across different CWEs. 
In summary, \tool contributes to more secure software development. 
Unlike existing code assistants, developers can integrate unlimited LLMs and code analysis engines into this framework through REST APIs.
\tool is open-source and publicly available on GitHub.\footnote{\url{https://github.com/aryakvnust/LLMSecGuard}}

The remainder of this paper is structured as follows.
In Section~\ref{motivation}, we motivate this work.
In Section~\ref{framework}, we introduce \tool, and in Section~\ref{usage}, we explain its two use cases.
In Section~\ref{relatedwork}, we present related work.
In Section~\ref{evaluation}, we outline our plans, and in Section~\ref{conclusion}, we conclude this paper.

\section{Motivation}
\label{motivation}

Security issues are pervasive in multiple domains~\cite{Buhlmann22}, from mobile applications~\cite{GhafariAndroidSmells2017, Gadient2019} and web servers~\cite{Gadient2021}, to critical software systems~\cite{Wetzels23}.
% Even worse, new technologies might bring back old vulnerabilities~\cite{Stivenart22}.
%
There are program analysis tools designed to uncover security issues, but studies indicate their limitations~\cite{Esposito2024,Corrodi2018}, as well as their lack of popularity among developers~\cite{Hazhirpasand2021}.
Unfortunately, the security landscape may not improve as we witness the popularity of language models for code generation~\cite{GitHubBlog}. It is concerning that developers place undue trust in these models, which are known to generate insecure code examples~\cite{wang2023effectiveness, Asare2023}.

In the following, we explain how we could iteratively ask ChatGPT 3.5 to generate secure code examples for symmetric encryption. 
Encryption plays a crucial role in our digital world, but its correct adoption has proven to be challenging for developers~\cite{Hazhirpasand18}.
For instance, the analysis of 489 open-source Java projects showed that 85\% of cryptography APIs are misused~\cite {Hazhirpasand2020}.
Importantly, existing tools that identify and fix such issues do not fulfill developers' expectations~\cite{Afrose2023, Zhang2023}.
%

%

% \newpage
Consider the StackOverflow post ID \href{https://stackoverflow.com/questions/3451670}{3451670},
% \footnote{\url{https://stackoverflow.com/questions/3451670}} 
which suffers from security issues such as ``ECB encryption mode'', ``static or constant key'', ``static salt for key derivation'', and ``hardcoded password''. 
When we presented this inquiry to ChatGPT, requesting a ``secure Java code example'', the resulting response still contained identical security issues.

\newpage
We then explicitly instructed ChatGPT that
\emph{
``I expect a secure encryption algorithm, a secure encryption mode, and a secure Initialization Vector (IV).
Do not use hard-coded secrets, suggest best practices, and make sure that your example provides confidentiality and integrity''}. 
Listing~\ref{lst:ChatGPTAnswer} presents the response.\footnote{https://chat.openai.com/share/8ce9eb6e-88b7-44dc-807a-51d04e7d446c}
% \href{https://chat.openai.com/share/8ce9eb6e-88b7-44dc-807a-51d04e7d446c}{response}.
% 

 \begin{lstlisting}[caption=ChatGPT's improved response to Post ID 3451670, label=lst:ChatGPTAnswer,float]
public class SecureEncryptionExample {
    public static void main(String[] args) throws Exception {
        String username = "bob@google.org";
        String password = "Password1";
        String secretID = "BlahBlahBlah";
        String SALT2 = "deliciously salty";
        // Generate a secure random IV (Initialization Vector)
        SecureRandom secureRandom = new SecureRandom();
        byte[] iv = new byte[12];
        secureRandom.nextBytes(iv);
        // Derive a key from the password using a secure method
        byte[] key = deriveKey(SALT2, username, password);
        // Create a SecretKey object using the derived key
        SecretKey secretKey = new SecretKeySpec(key, "AES");
        // Instantiate the cipher with AES in GCM mode
        Cipher cipher = Cipher.getInstance("AES/GCM/NoPadding");
        ...
    }
    private static byte[] deriveKey(String salt, String username, String password) throws Exception {
        // Concatenate salt, username, and password
        String combined = salt + username + password;
        // Use a secure hash function (SHA-256) to derive a key
        MessageDigest md = MessageDigest.getInstance("SHA-256");
        byte[] hash = md.digest(combined.getBytes());
        return Arrays.copyOf(hash, 16);
    }
  } 
\end{lstlisting}

ChatGPT identified the constant key, and it correctly suggested the secure ``GCM'' encryption mode as a best practice. However, violations such as ``static salt for key generation'' and ``hardcoded password'' were in the response without any proper warning. 
The suggested code, at line 23, included the ``SHA-256'' hash function for key generation, which is not a best practice.
Anyway, 
when we asked about the security risk at line 6 (ie., static salt), ChatGPT responded: 
\emph{``Using a hard-coded string as a salt is generally not considered a good practice for security.
Ideally, salt should be unique for each user and should be randomly generated for each password. Using a constant salt for all users could make the system more vulnerable to certain types of attacks, such as rainbow table attacks''}. 
It then provided the code example in Listing \ref{lst:ChatGPTAnswer2} as a replacement.

 \begin{lstlisting}[caption=ChatGPT's suggestion to replace a static salt, label=lst:ChatGPTAnswer2]
// Generate a secure random salt
SecureRandom secureRandom = new SecureRandom();
byte[] salt = new byte[16];
secureRandom.nextBytes(salt);
// Convert the salt to a Base64-encoded string for storage
String SALT2 = Base64.getEncoder().encodeToString(salt);

\end{lstlisting}

Subsequently, we asked, \emph{``Are there any other security risks in your provided code?''}. 
It failed to detect the hardcoded password, but at this stage, it correctly suggested a key derivation function (KDF) instead of the hash function for generating a password-based key.
Upon a direct inquiry about the line where a hardcoded password existed, it offered a secure suggestion.

\boxit[yellow!20]{
Through proper interactions with ChatGPT, specifically by listing potential security issues and their locations within the code, it is feasible to harness ChatGPT's power to generate secure code examples.
However, novice developers may not possess the necessary expertise to provide such inputs. 
Hence,
we introduce \tool, a framework that links LLMs and static analysis tools to overcome this limitation. 
}

\section{\NoCaseChange{\tool}}
\label{framework}

\begin{figure}[!h]
	\centering
	\includegraphics[scale=0.80]
 % [width=1\columnwidth]
 {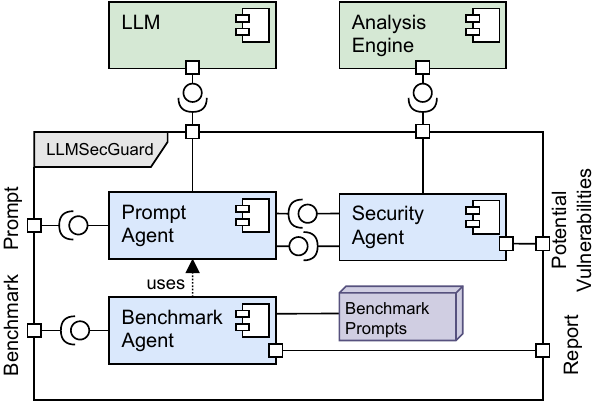}
	\caption{\tool's components}
	\label{fig:arch}
\end{figure}

We introduce \tool, an open-source framework to offer enhanced code security through the synergy between code security analyzers and LLMs.
The primary objective of \tool is (i) to enhance the security of LLM-generated code, and (ii) to benchmark the security properties of LLMs. 
It adopts a RESTful architecture, implemented in Python using   
Django and Flask, and uses SQLite for data persistence.
% \tool is publicly available on GitHub.\footnote{\url{https://github.com/aryakvnust/LLMSecGuard}}
Figure~\ref{fig:arch} illustrates the schematic architecture of \tool. We explain the main components in the following. 

\emph{Prompt Agent.} 
This component is responsible for receiving a prompt and providing other components with the code model's response. 
Upon receipt of a prompt, ``Prompt Agent'' undertakes the task of engineering a response.
Particularly, it can reformulate a prompt, pass the prompt to code models, collect the response, and forward the result to other components.

\emph{Security Agent.} 
This component has a pivotal role in uncovering security issues in LLM-generated code.
Precisely, it is entrusted to pass the code to static code analysis engines (such as Semgrep and Weggli), and to collect potential security vulnerabilities.

\emph{Benchmark Agent.}
This component puts different LLMs to security test. 
Particularly, it evaluates the security properties of LLMs based on a set of challenges structured in JSON format. 
Each challenge includes a prompt and the expected outcome, and ``Benchmark Agent'' is responsible for flagging LLMs that pass the challenge.

\section{Usage Scenarios}
\label{usage}

% We explain the operational workflow of \tool.
We describe the \tool's main configurations including its connection with external entities such as LLMs and code analyzers.
We then explain the interplay between different components through two main usage scenarios, namely ``benchmarking'' and ``code generation''.

% The tool is primarily designed for developers who are interested in discerning the reliability of various AI code assistants, and those seeking enhanced security solutions.

\subsection{Setup}
There are several key configurations for running \tool that users can adjust according to their preferences.

\tool requires a minimum of one LLM and one code analysis engine to operate effectively. This can be easily set up by providing the API information (e.g., API endpoint and API key) associated with each entity. There is no limitation, and users can add as many entities as they wish. 
Currently, we have instantiated \tool with ANYSCALE Llama2~\cite{MetaAILLAMA}, and Weggli and Semgrep static code security 
analysis tools~\cite{SemgrepRepo, WeggliRepo}.
Weggli is a robust and efficient semantic code analysis for C and C++ programs, and Semgrep is designed for programs written in languages such as C\#, Java, Java, JavaScript, Python, PHP, Ruby, Scala, etc.

\tool relies on CyberSecEval, a benchmark specifically designed to evaluate the cybersecurity aspects of LLMs functioning as coding assistants~\cite{bhatt2023purple}. 
It comprises exhaustive test cases to assess the potential of LLMs in producing insecure code and facilitating cyberattacks. 
Nonetheless, 
users have the flexibility to update this benchmark or replace it with one of their preferences.

\tool includes a few execution parameters as well. 
The primary one is the ``termination condition'' designed to prevent the system from looping indefinitely.
This condition determines how many iterations a code piece should undergo improvement and analysis if potential vulnerabilities persist. 

The ``benchmark interval'' parameter determines how often the benchmark process is executed, with options such as biweekly or monthly intervals. Given that LLMs evolve and their security properties may change, regular updates are necessary to ensure that results remain current.

Lastly, 
users can choose the LLMs of interest for benchmarking, select specific code analysis engines, and specify challenges that should be considered for measurement.

\subsection{Security Benchmarking}

In this scenario, a set of existing prompts, referred to as benchmark prompts, undergo evaluation.
In particular, 
the ``Benchmark Agent'' sends each prompt in the benchmark to the ``Prompt Agent''. 
This component dispatches the prompt to every LLM considered for benchmarking, collects responses for each prompt, and forwards them to the ``Benchmark Agent''.
Each prompt presents a challenge, and ``Benchmark Agent'' determines the extent to which an LLM succeeds in resolving it by comparing the response to the expected outcome.
For instance, in terms of vulnerable code, it measures the quantity and severity of potential CWEs.
Each LLM is assigned a performance score for each challenge, and ultimately, LLMs are ranked.

% The resulting data is then stored in a database for subsequent analyses, such as ranking LLMs across different challenges.

%
At present, \tool relies on the CyberSecEval benchmark, which incorporates extensive prompts and regular expressions to uncover potential issues. 

\subsection{Secure Code Generation}
\begin{figure}[!h]
	\centering
	\includegraphics[scale=0.75]
 {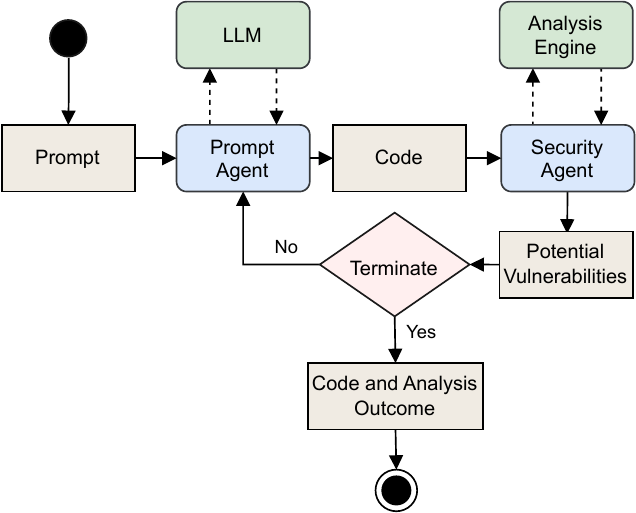}
	\caption{\tool's secure code generation workflow}
	\label{fig:workflow}
\end{figure}

Figure~\ref{fig:workflow} illustrates the workflow in this scenario.
It begins with the user providing a prompt of interest, anticipating the output to be a piece of code.
The ``Prompt Agent'' assumes control and forwards the user's prompt to the top LLM determined in the benchmarking scenario (unless the user chooses a different LLM).
Upon receiving the LLM's response, this component then transfers the produced code to the ``Security Agent'' for security inspection.
The ``Security Agent'' collaborates with external analysis engines (configured earlier) to uncover potential vulnerabilities and respective lines in the code.
If the termination condition is met (i.e., there is no vulnerability or the maximum analysis iterations is reached), the code, along with its vulnerability information, is immediately reported back to the user.
In cases where the termination condition is not met,
this information is relayed to the ``Prompt Agent''.
This component formulates a new prompt based on the collected vulnerability information and queries the LLM once again.

\section{Relate Work}
\label{relatedwork}

The popularity of AI-generated code has attracted the attention of security researchers.
Pearce et al. assessed Copilot's performance in suggesting code across 89 scenarios aligned with MITRE's ``Top 25'' CWEs, revealing that approximately 40\% of the generated programs contained vulnerabilities~\cite{pearce2022asleep}.
Nonetheless, in a later study, Asare et al. came to a different conclusion~\cite{Asare2023}. 
Precisely, Copilot did not consistently reproduce past vulnerabilities introduced by humans. 
In about 25\% of cases, Copilot even proposed fixes for these vulnerabilities, suggesting a potentially lower susceptibility to security risks compared to human programmers.
Fu et al. analyzed 435 code snippets generated by Copilot in public GitHub projects and found that 35\% of the snippets exhibit CWEs~\cite{fu2023security}.
Mousavi et al. designed 48 programming tasks for five common security APIs in Java and evaluated the performance of ChatGPT in developing these tasks. They uncovered a concerning level of security, specifically, they found an average misuse rate of 70\% in these tasks~\cite{mousavi2024investigation}.

Researchers have also developed benchmarks for assessing the security of LLM-generated code.
Bhatt et al. developed CYBERSECEVAL, a benchmark tailored to evaluate the cybersecurity risks posed by LLMs~\cite{bhatt2023purple}.
Hajipour et al. examined code models for generating high-risk security weaknesses and built a collection of diverse non-secure prompts for various vulnerability scenarios, which can be used as a benchmark to compare security weaknesses in LLMs~\cite{Hajipour2024}.

\section{Future Work}
\label{evaluation}

We plan to investigate whether \tool will effectively support developers in writing secure code in real-world scenarios. Two groups of developers, both utilizing LLMs for coding, will be recruited for our study. 
We will assign the same programming tasks to each group, instructing one to utilize LLMs freely and the other to exclusively use \tool during the coding process.
We will measure the time taken to complete each task, the number, and the severity of vulnerabilities.
We will compare the results across two groups as well as based on participants' experience. 

Pending positive evaluations, our ultimate goal is to integrate \tool into at least one popular IDE (Integrated Development Environment), as deemed essential by developers for a seamless user experience~\cite{Tymchuk2018}. 
Furthermore, this integration would allow \tool to gather development context, thereby enabling the formulation of better prompts for code generation.

Future work could also investigate code changes in each iteration between ``Prompt Agent'' and ``Security Agent'', as well as examine how the engineering of prompts and vulnerability information affects the performance of LLMs.

Finally, it is noteworthy that although a code snippet may not inherently contain vulnerabilities, its integration into a program or execution within a specific environment could still pose security risks and therefore warrants investigation~\cite{Stivenart22}.

\section{Conclusion}
\label{conclusion}

We introduced \tool, an open-source framework developed to equip developers with code solutions that are more secure than the code initially generated by Large Language Models (LLMs). These code suggestions are obtained through the integration of LLMs and static security code analyzers. \tool also offers a feature to measure the security properties of LLMs and provide a current security comparison of different LLMs in the wild.

\bibliographystyle{ACM-Reference-Format}
\bibliography{ease2024-81}

\end{document}